\documentclass{ws-ijmpa}
\usepackage[super,compress]{cite}
\usepackage{graphicx}
\usepackage{subfigure}
\usepackage{float}
\usepackage{xcolor}
\usepackage{multirow}
\usepackage{lscape}
\usepackage{tensor}
\usepackage{hyperref}
\usepackage{lscape}
\begin{document}
\markboth{Carrillo-Monteverde et.al}{On the Universal Texture in the
  PA-2HDM for the V-SPIN case} \bibliographystyle{ws-ijmpa}
%
\catchline{}{}{}{}{}
%
\title{\bf On the Universal Texture in the PA-2HDM for the V-SPIN
  case}

\author{A. Carrillo-Monteverde\footnote{alba\_carrillo@uaeh.edu.mx} ,
  S. G\'omez-\'Avila\footnote{selim\_gomez@uaeh.edu.mx} ,
  L. L\'opez-Lozano\footnote{lao\_lopez@uaeh.edu.mx}
  \footnote{Corresponding author.}}

\address{\'Area Acad\'emica de Matem\'aticas y F\'isica,
  Universidad Aut\'onoma del Estado de Hidalgo,\\
  Carr. Pachuca-Tulancingo Km. 4.5, C.P. 42184, Pachuca, Hidalgo,
  Mexico.  }

\maketitle

\begin{history}
  \received{Day Month Year} \revised{Day Month Year}
\end{history}

\begin{abstract}
  In a Partially Aligned Two Higgs Doublet Model, where only is
  allowed flavor violation between third and second generation of
  fermions, we propose a mechanism to generate the second Yukawa
  matrix through a Unitary V-Spin flavor transformation on the mass
  matrix for quarks and leptons. Also we assume that this flavor
  transformation is universal, this is, we use the same parameters to
  generate Yukawa matrix elements in both sectors, reducing
  drastically the number of free parameters. As consequence we
  obtained a serie of relations between Yukawa matrix elements, that we called
  the Universality Constraint.  Also, we obtained an interval of values for
  the second Yukawa matrix elements, expressed in terms of the Cheng
  and Sher ansatz, for $\tau\to\mu\mu^+\mu^-$ and $\tau\to\gamma\mu$
  coming from the Universality Constraint and experimental bounds for
  light scalar masses. Finally, we show the allowed region of
  parameters for the flavor transformation from $B_s\to\mu\mu$ decays,
  $B_s^0-\bar{B}_s^0$ mixing, $\tau\to\mu\mu^+\mu^-$ and
  $\tau\to\gamma\mu$ experimental bounds.

\end{abstract}

\ccode{PACS numbers:11.30.Hv, 12.15.Ff, 13.20.-v, 14.80Cp }

\maketitle
\section{Introduction}

After the 2012 discovery of the Higgs boson, particle physics entered
a new era. On one hand, the Standard Model (SM) predictions have been
confirmed with remarkable precision, reinforcing its role as the best
description of nature up to the electroweak scale. On the other hand,
there are many open questions from the theoretical and
phenomenological points of view, and some tension in the experimental
measurements related to some rare meson decays. At present, we do not
have a deep understanding of the matter sector of the SM,
its familiar multiplicity and hierarchical couplings, nor can we
explain the amount of baryonic asymmetry and the abundance and nature
of dark matter. Beyond the SM, theories with Natural
Flavor Conservation are unable to account for various tensions between
the SM and Lepton Flavor Violation (LFV), polarization asymmetries,
etc.  At the same time, many observables constrain the apparition of
Flavor Changing Neutral Currents (FCNCs), which should be very
small\cite{Tanabashi2018}.

An interesting feature of the SM as it stands today is
that the mixing structures of quarks and leptons are notably
different: neutrino mixing is very large compared with quark mixing,
where the offline CKM entries are significantly supressed. This
appears to suggest that different mechanisms may account for lepton
and quarks mass generation and mixings. For example, this happens in models
that couple the light and heavy fermions to different Higgs
doublets\cite{Altmannshofer2018}. Another interesting possibility is
the existence of Majorana mass terms in the neutrino sector, which
would provide a clear difference between quarks and
leptons\cite{Branco2015}.

However, apparently different mixing structures in the leptonic and quark sectors, can be accomodated with the same
structure at the price of leaving the fermion masses unexplained. This
is the idea of a universal texture \cite{Koide2004, Matsuda2004},
where the mass hierarchy is responsible for the nearly maximal
neutrino mixing and the suppressed quark mixing angles. In this
scenario, only the mixing is addressed since the lepton and quark
mass hierarchy remains unexplained, although it is possible to
implement such hierarchy with a Froggat-Nielsen mechanism
\cite{Froggatt:1978nt}. Nonetheless, while representing only a partial
solution, this would still be a step forward.

The present work explores this concept by pitting a version of the
universal texture idea against the significant constraints coming from
rare lepton and meson decays. In a previous
work\cite{Carrillo-Monteverde2019}, a version of the 2HDM was studied
where Yukawa matrices are almost-aligned in such a way that only a
pair of generations at a time develops FCNC's. We will refer to this
model as the Partially Aligned Two Higgs Doublet Model (PA-2HDM)
\cite{Hernandez-Sanchez2012a}, and focus in an scenario with second
and third family mixing, called the V-spin. In order to implement the
universal texture idea, the same dimensionless parameters are used for
the quark and lepton Yukawa couplings, with scales provided by the
fermion masses.

In the following, section \ref{sec:vspin} lays out the model and the
assumptions employed. In section \ref{sec:rare}, some rare decays are
used to test the viability of the model and we show our numerical
analysis. Afterwards, we present our conclusions in section
\ref{sec:conc}.

\section{The V-Spin case of the PA-2HDM and the Universality Constraint}\label{sec:vspin}
The starting point of the PA-2HDM is a generic 2HDM that might be seen
as a low energy limit of a model with an underlying flavor
dynamics. The Yukawa sector of the 2HDM has the following form
\begin{eqnarray}\label{lag_PA2HDM}
  -\mathcal{L}_Y&=&\bar{Q}_{Li}\left[Y^d_{a,ij}\Phi_a D_{Rj}+Y^u_{a,ij}\tilde{\Phi}_aU_{Rj}\right]\nonumber\\
                &&+\bar{L}_{Li}Y_{a,ij}^\ell\Phi_a E_{Rj}+\text{H.c.},
\end{eqnarray}
where $i,j$ are flavor indexes and $a=1,2$. $Q_{Li}= (U_{Li},D_{Li})$
and $L_{Li}= (N_{Li},E_{Li})$ are the fermion $SU(2)$ doublets, and
$D_{Rj}$, $U_{Rj}$ and $E_{Rj}$ are the fermion singlets. The mass
matrix is defined by
\begin{equation}
  \label{eq:mass_matrix}
  M^f_{ij}=\frac{1}{\sqrt{2}}\left(v_1Y^f_{1,ij}+v_2Y^f_{2,ij}\right),
\end{equation}
where $f=u,d,\ell$ or $e,\mu,\tau$, and $v_1$ and $v_2$ are the vacuum
expectation values (VEV) of each doublet. There are several ways to
parameterize flavor violating (FV) couplings. In this work, to ease
comparison with the literature, we use (\ref{eq:mass_matrix}) to write
down the Feynman rules in terms of the mass matrix $M^f_{ij}$, and
$Y^f_{2,ij}$ for the couplings of down-type quarks and leptons. Once
the mass matrix is diagonal in the mass basis, the matrix $Y^f_{2,ij}$
contains all information of the FV contribution.

We express the Yukawa matrix elements in the Cheng and Sher ansatz
\cite{Cheng:1987rs}, but relaxing the requirement of order-one
couplings $\chi^f_{2,ij}$:
\begin{equation}
  \label{eq:cheng_and_sher}
  Y^f_{2,ij}=\frac{\sqrt{m^f_im^f_j}}{v}\chi^f_{2,ij}, \quad\text{for} \quad f=d,\ell .
\end{equation}
Here, the parameters $\chi^f_{2,ij}$ can take any complex value.
Without loss of generality, we can parameterize these contributions as
follows:
\begin{equation}
  \label{eq:flavor_transformation}
  \chi_{2,ij}^f= \frac{1}{\sqrt{m^f_im^f_j}}\left(A^{f}_{2L}M^f A_{2R}^{f}\right)_{ij},
\end{equation}
where $v^2=v_1^2+v_2^2$ and $A_{2L(R)}^f$ are $SU(3)$ matrices.

Alignment between the Yukawa couplings\cite{Pich:2009sp} is one of
several ways to constrain the large number of parameters of the
2HDM-III\cite{Atwood:1996vj}. As it was described before, the PA-2HDM
relaxes the alignment in order to generalize the model while keeping
the free parameters manageable. In this work, we will focus in a
version of the PA-2HDM where only the mixing between the second and
the third generation of quarks and leptons is allowed. We consider
this to be the most promising case given the experimental
constraints. For example, the $K^0-\bar{K}^0$ mixing is very close to
the SM prediction, and therefore permits a very restricted NP
contribution \cite{Antaramian1992}. In the leptonic $\tau$ decays and $\mu$ decays
recently updated upper bounds are given by
\cite{Tanabashi2018,Dib2019a}

\begin{eqnarray}
  \text{Br}(\mu\to ee^+e^-)&<1.0\times 10^{-12},\\
  \text{Br}(\mu\to e\gamma\gamma)&<1.0\times 10^{-12},\\
  \text{Br}(\tau\to \mu e^+e^-)&<1.8\times 10^{-8},\\
  \text{Br}(\tau\to \mu \gamma)&<4.4\times 10^{-8},\\
  \text{Br}(\mu\to e \gamma)&<4.2\times 10^{-13}.
\end{eqnarray}

The flavor changing (FC) upper bounds for $\tau$-$\mu$ are up to five
orders of magnitude larger than $\mu$-$e$ bounds. If we include the FC
processes $\tau$-$e$, we have bounds similar to $\tau$-$\mu$. Modulo
kinematical effects, the mixing of any generation with the third
generation prevails over the one that first and second mixing. Now, if
we observe the leptonic decays\cite{Tanabashi2018}
\begin{eqnarray}
  \text{Br}(B_s^0\to\mu\bar{\mu})=(3.0\pm 0.4)\times 10^{-9},\\
  \text{Br}(B_d^0\to\mu\bar{\mu})=(1.4^{+1.6}_{-1.4})\times 10^{-10},
\end{eqnarray}
the FV process for $b-s$ mixing is one order of magnitude larger than
the $b-d$ one. Since we want to find a common mixing texture for
leptons and quarks, this favors choosing the PA-2HDM version with
mixing of the second and third generations. We call the choice to
enforce a universal texture the Universality Constraint(UC). Having
taken observations on the current experimental evidence, and
considering that we are looking for a scenario with light pseudoscalar
and scalars, we will from now on consider the PA-2HDM with
$\chi^{q,\ell}_{2,1j}\simeq 0$ and $j=(2,3)$, the V-Spin scenario.

In order to analyze the consequences of the UC and its
phenomenological viability, we have chosen a set of channels where
FCNC and LFV couplings appear either at tree level (in the case of
$B_s^0\to\mu\bar{\mu}$, $B^0_s-\bar{B}^0_s$ physics), or at least at
one and two loops (for $\tau\to\mu\ \mu^+\mu^-$,
$\tau\to\gamma\mu$). In Table \ref{channels}, we show the experimental
values for the branching ratios and the SM bounds.  The $\tau$ channels
are promising since the corresponding SM contribution is extremely small.

\begin{center}
  \begin{table}[hp]\label{channelsVSpin}
    \tbl{Channels with Flavor Violation where only couplings of the
      second and third generation $\chi_{2,ij}^{q,\ell}$ are
      involved.}  {
      \begin{tabular}{cll}
	\hline
	Channel & SM prediction & Exp. Results [\citen{Tanabashi2018}]\\
	\hline
	$\text{BR}(\tau \to\gamma\mu)$& $ <10^{-53}$ [\citen{Lee1977, Kou2019}]& $<4.4\times 10^{-8}$\\
	$\text{BR}(\tau\to \mu \mu^+\mu^-)$ &$6.4\times 10^{-55}$ [\citen{Hernandez-Tome2019}]&$<2.1\times 10^{-8}$\\
	$\text{BR}(B_s^0\to\mu^+\mu^-)$ &$(3.65\pm0.23)\times 10^{-9}$ [\citen{Bobeth2014d}]&$(3.0\pm 0.4)\times10^{-9}$\\
	$\Delta M_{B^0_s}\  (\times 10^{-8} \text{MeV})^{-1}$&$(1.3167\pm 0.08225)$ [\citen{DiLuzio2018}]&$(1.169\pm 0.0014)$\\
	\hline
      \end{tabular}
      \label{channels}
    }
  \end{table}
\end{center}

Because we have chosen the V-Spin scenario, as discussed in a previous
work\cite{Carrillo-Monteverde2019}, channels with FCNC and LFV
involving fermions of the first generation do not get NP
contributions. Therefore, the compatibility between the SM predictions
and the experimental data will not be altered. Also, the values for
flavor-conserving parameters $\chi^{\ell,q}_{2,ii}$ are not strongly
constrained.

The observations described above are similar to those that sustain the
approximations taken long time ago in the context of the 2HDM-III as
in reference [\citen{Xiao:2003ya}], or even to ignore the tree level
FC coupling in [\citen{BowserChao:1998yp}], to mention
some examples. In contrast, in this work we assume that the absencet
of FC couplings at tree level involving the first
generation of fermions is consequence of an underlying flavor symmetry
in the context of a generic 2HDM.

For the V-Spin texture, we can rewrite the couplings in terms of the
parameters $(c^V_0, c^V_3, c^V)$ and the angle $\theta_V$:
\begin{eqnarray}
  \tilde\chi^{f}_{2,11}&=&(c^V_0)^2, \label{eq:VSpinChi0}\\
  \tilde\chi^{f}_{2,22}&=&(c^V_0+ c^V_3)^2+\frac{m^f_3}{m_2^f}(c^V)^2, \\
  \tilde\chi^{f}_{2,33}&=&(c^V_0- c^V_3)^2+\frac{m^f_2}{m_3^f}(c^V)^2, \\
  \tilde\chi^{f}_{2,23}&=&c^Ve^{i\theta_V}\left[\sqrt{\frac{m^f_2}{m_3^f}}(c^V_0+c^V_3) + \sqrt{\frac{m^f_3}{m^f_2}}(c^V_0-c^V_3)\right],\label{eq:VSpinChi}
\end{eqnarray}
with $\tilde\chi_{2,13}^{f}=\tilde\chi_{2,12}^{f}=0$.

This same structure, with the same values for the $c^V$, $c_0^V$ and
$c_3^V$ parameters, is used for the leptonic and quark sectors.

\section{Rare $\tau$ decays and FCNC for $B_s^0$ in the V-Spin
  texture}\label{sec:rare}

\subsection{The observables for the V-Spin texture}
In the PA-2HDM for the V-Spin case, the only channels with LFV are those that involve the $\tau$ and $\mu$
leptons. We will consider the decays of $\tau$ leptons that appear in
Table \ref{channels} in order to test our hypothesis about the
viability of the reduction of free parameters.

\subsubsection{The  $\tau \rightarrow \mu \gamma$ channel}
For $\tau \rightarrow \mu \gamma$, the SM prediction is too small to
be experimentally accesible\cite{Hernandez-Tome2019}, while the
experimental bound is many orders of magnitude
above\cite{Aubert2010}. This bound is expected to improve one order of
magnitude during the current LHC era\cite{Aushev2010}. The
Bjorken-Weinberg mechanism can make the two loops correction larger
than the one loop contribution\cite{Bjorken1977}. This effect is more
pronounced in the higher mass region. Here, we are mainly interested
in relatively light new scalars from the 2HDM, where the two-loops
contribution is not too important, nevertheless we will consider it in
our numerical calculations. Also, as this decay has been measured with
very good precision, we use the exact 2-loops expression given
by\cite{Chang1993,Davidson2010}
\begin{eqnarray}
  \mathcal{M}&\simeq & \frac{1}{16\pi^2}\left\{ \sqrt{2}\sum_{\phi } \frac{\Gamma_{\phi\mu\tau}\Gamma_{\phi\tau\tau}}{M_\phi ^2} \left(\log\frac{M_\phi^2}{m_\tau^2}-\frac{3}{2}\right)\right.\nonumber\\
             && \quad\left. +2\sum_{\phi,f}\Gamma_{\phi\mu\tau}\Gamma_{\phi f f}\frac{N_cQ_f^2\alpha}{\pi}\frac{1}{m_\tau m_f}F_\phi\left(\frac{m_f^2}{M_\phi^2}\right)\right.\nonumber\\
             &&\left.\qquad-\sum_{h^0,H^0}\Gamma_{\phi\mu\tau}C_{\phi W W}\frac{g \alpha}{2\pi m_\tau M_W}\left[3F_\phi\left(\frac{M_W^2}{m_\phi^2}\right)\right.\right.\nonumber\\
             &&\qquad\left.\left. +\frac{23}{4}G\left(\frac{M_W^2}{M_\phi^2}\right)+\frac{3}{4}H\left(\frac{M_W^2}{M_\phi^2}\right)+M_\phi^2\frac{F_\phi\left(\frac{M_W^2}{M_\phi^2}\right)-G\left(\frac{M_W^2}{M_\phi^2}\right)}{2M_W^2}\right]\right\}
\end{eqnarray}

where $\phi =h^0,H^0,A^0$ and $f=t,b$ . The loops integrals can be
written as\cite{Chang1993}
\begin{eqnarray}
  F_A(z)&=&G(z)=\frac{z}{2}\int^1_0 dx \frac{1}{x(1-x)-z}\log\left[\frac{x(1-x)}{z}\right],\\
  F_{h^0,H^0}&=&\frac{z}{2}\int^1_0 dx \frac{(1-2x(1-x))}{x(1-x)-z}\log\left[\frac{x(1-x)}{z}\right],\\
  H(z)&=&-\frac{z}{2}\int^1_0dx \frac{1}{x(1-x)-z}\left\{1-\frac{z}{x(1-x)-z}\log\left[\frac{x(1-x)}{z}\right]\right\}.
\end{eqnarray}

In these expressions, only the 2-loop contributions coming from
virtual $t$ and $b$ quarks have been considered. 

The couplings for the PA-2HDM are obtained from
(\ref{eq:cheng_and_sher}) and (\ref{eq:VSpinChi}). In this model, the
couplings of gauge bosons with the scalars $H^0$ and $h^0$ are the
same as in Ref. [\citen{Davidson2010}]:
\begin{eqnarray}
  C_{h^0 W W}&=&\sin(\beta-\alpha),\\
  C_{H^0 WW}&=&\cos(\beta-\alpha).
\end{eqnarray}

\begin{figure}[ht]\label{tautogmu}
  \begin{center}
    \includegraphics[scale=0.60]{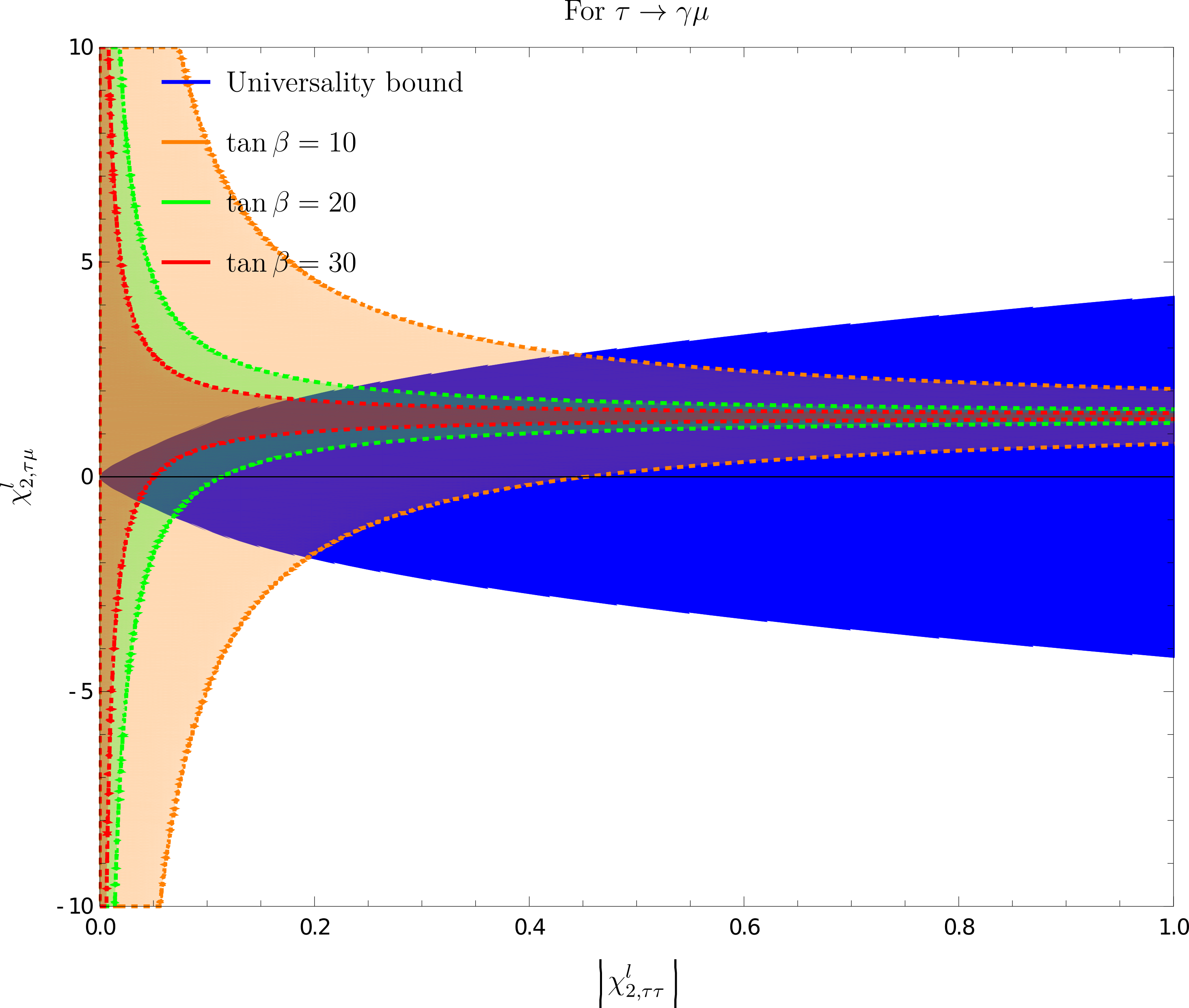}
    \caption{Alowed regions for $\chi^\ell_{2,\tau\mu}$ and
      $|\chi^\ell_{2,\tau\tau}|$ with masses $m_h=125.6$ GeV,
      $M_{H^0}=300$ GeV and $M_{A^0}=300$ GeV and several values of
      $\tan\beta$. Constrictions come from the $\tau\to\gamma\mu$
      experimental bound and from the UC hypothesis. }
  \end{center}
\end{figure}

The branching ratio for this decay is a function of the scalar masses,
$\tan\beta$, the angle $\alpha$ and the $c's$ parameters through the
$\chi^\ell_{2,\tau\mu}$ and $\chi^\ell_{2,\tau\tau}$ couplings. In
order to test the V-Spin texture and compute our numerical estimations
we have fixed $\alpha=0$ and set relatively light scalar masses;
\emph{i. e.}  $M_{H^0}=300$ GeV and $M_{A^0}=300$. For the light
scalar $h^0$, we have fixed its mass as $125.6$ GeV. In Figure
\ref{tautogmu}, we show the allowed regions in the parameter space
$\chi^\ell_{2,\tau\mu}$ \emph{vs} $\chi^\ell_{2,\tau\tau}$ that are
compatible with the experimental upper bound of
$\text{Br}(\tau\to\gamma\mu)$ for different values of
$\tan\beta$. Also, we plot the restricted region that comes from the
UC, that is, the restriction coming from the expressions
(\ref{eq:VSpinChi0}-\ref{eq:VSpinChi}). As it can be seen in the
graph, for small values of $|\chi^\ell_{2,\tau\tau}|$ the upper bound
of $\chi^\ell_{2,\tau\mu}$ is determined by the UC and the allowed
region is smaller than the one without V-Spin texture.

\subsubsection{The channel $\tau\to 3\mu$}
\begin{figure}[ht]\label{tauto3mu}
  \begin{center}
    \includegraphics[scale=.60]{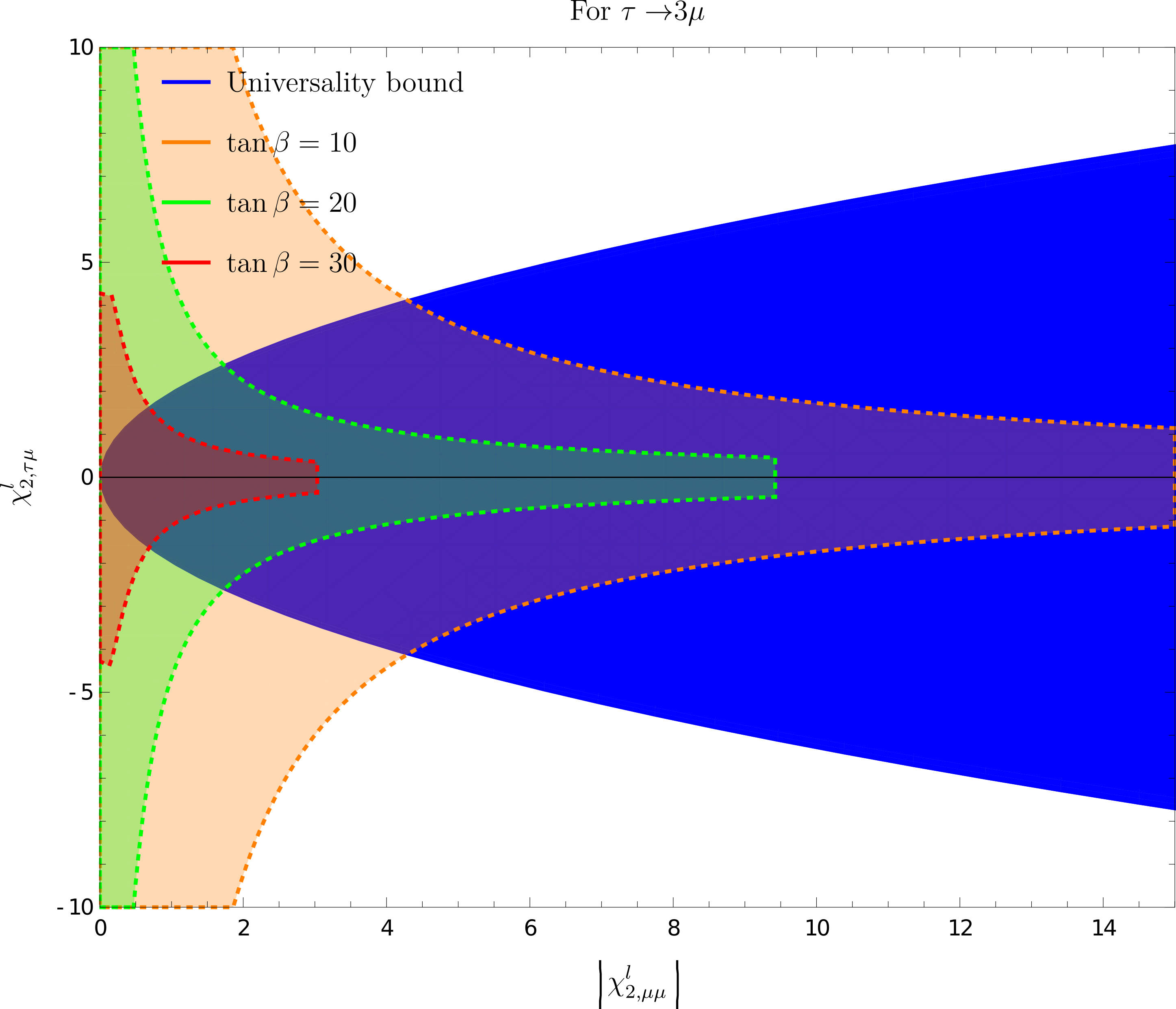}
    \caption{Here, it is shown the allowed values for
      $\chi^\ell_{2,\tau\mu}$ and $|\chi^\ell_{2,\mu\mu}|$ for
      different values of $\tan\beta$ and masses $M_{h^0}=125.6$ GeV,
      $M_{H^0}=300$ GeV and $M_{A^0}=300$ GeV. Also it is shown the
      UC coming from the assumption of a common
      V-Spin texture. As we can see, for small values of
      $|\chi^\ell_{2,\mu\mu}|$ the main restriction on
      $\chi^\ell_{2,\tau\mu}$ comes from the UC
      and there is a significant reduction of the allowed region.}
  \end{center}
\end{figure}

In the SM, the $\tau \to\mu\mu^+\mu^-$ decay is strongly suppressed
because the main contribution comes from higher order diagrams
suppressed by a GIM-like mechanism\cite{Hernandez-Tome2019}. In order
to estimate the NP contribution to this channel, we dismiss the SM
contribution and adopt the formalism of the Effective Lagrangians
\cite{Celis2014,Dib2019}. In general, the amplitude for this channel
from scalar contributions can be written as
\begin{eqnarray}
  \mathcal{L}_\text{eff}&\simeq& C_{\tau\mu\mu\mu}^{LL}\left(\bar\mu P_L \tau\right)\left(\bar\mu P_L\mu\right)+C_{\tau\mu\mu\mu}^{RR}\left(\bar\mu P_R \tau\right)\left(\bar\mu P_R\mu\right)\nonumber\\
                        &&+C_{\tau\mu\mu\mu}^{LR}\left(\bar\mu P_L \tau\right)\left(\bar\mu P_R\mu\right)+C_{\tau\mu\mu\mu}^{RL}\left(\bar\mu P_R \tau\right)\left(\bar\mu P_L\mu\right)
\end{eqnarray}
where the Wilson coefficients are defined by
\begin{equation}
  C_{\tau\mu\mu\mu}^{Q_1Q_2}=-i\sum_{\phi}\frac{\Gamma_{\phi \tau\mu}^{Q_1}\Gamma_{\phi \mu\mu}^{Q_2}}{M_\phi^2}
\end{equation}
with $Q_1=L,R$ and $Q_2=L,R$ and $\phi=h^0,H^0,A^0$. The couplings
obtained from the Lagrangian (\ref{lag_PA2HDM}) using the
parametrization described in section \ref{sec:vspin} are
\begin{eqnarray}
  \Gamma^{R(L)}_{h^0 \ell_i\ell_j}&=&+(-)i\left[\frac{\cos(\alpha-\beta)}{v\cos\beta}\delta_{ij}m_{\ell_i}-\frac{\sin\alpha}{\sqrt{2}\cos\beta}Y^\ell_{2,ij}(c^V,c_0^V,c_3^V)\right]\\
  \Gamma^{R(L)}_{H^0 \ell_i\ell_j}&=&+(-)i\left[\frac{\sin(\alpha-\beta)}{v\cos\beta}\delta_{ij}m_{\ell_i}+\frac{\cos\alpha}{\sqrt{2}\cos\beta}Y^\ell_{2,ij}(c^V,c_0^V,c_3^V)\right]\\
  \Gamma^{R(L)}_{A^0 \ell_i\ell_j}&=&-\left[\frac{\tan\beta}{v\cos\beta}\delta_{ij}m_{\ell_i}-\frac{1}{\sqrt{2}\cos\beta}Y^\ell_{2,ij}(c^V,c_0^V,c_3^V)\right]\label{couplingsGammaA}
\end{eqnarray}
where the matrices $Y^\ell_{2,ij}$ are hermitian. Thus, the squared
amplitude takes the following form
\begin{eqnarray}\label{amplitude2}
  |\mathcal M|^2&=&4\left[\frac{1}{4}(m^2_{23}-2m_{\mu}^2)(m_\tau^2+m_{\mu}^2-m_{23}^2)F_1+m_\tau m_\mu(m_{23}^2-2m_\mu^2)F_2\right.\nonumber\\
                &&+m_\mu^2(-m_{23}^2+m_\tau^2+m_\mu^2)F_3+ 2m_\tau m_\mu^3F_4\Big]
\end{eqnarray}
where $m_{23}^2=(p_2+p_3)^2$ and $p_{2,3}$ are the 4-momentum of the
final muons. The parameters $F_i$ are in general
\begin{eqnarray}
  F_1&=&|C^{LL}_{\tau\mu\mu\mu}|^2+|C^{LR}_{\tau\mu\mu\mu}|^2+|C^{LR}_{\tau\mu\mu\mu}|^2+|C^{RL}_{\tau\mu\mu\mu}|^2,\\
  F_2&=&\text{Re}\left(C^{RL}_{\tau\mu\mu\mu}C^{LL*}_{\tau\mu\mu\mu}+C^{RR}_{\tau\mu\mu\mu}C^{LR*}_{\tau\mu\mu\mu}\right),\\
  F_3&=&\text{Re}\left(C^{LR}_{\tau\mu\mu\mu}C^{LL*}_{\tau\mu\mu\mu}+C^{RR}_{\tau\mu\mu\mu}C^{RL*}_{\tau\mu\mu\mu}\right),\\
  F_4&=&\text{Re}\left(C^{RR}_{\tau\mu\mu\mu}C^{LL*}_{\tau\mu\mu\mu}+C^{RL}_{\tau\mu\mu\mu}C^{LR*}_{\tau\mu\mu\mu}\right).
\end{eqnarray}
With the amplitude (\ref{amplitude2}), we have calculated the
branching ratio of the $\tau\to 3\mu$ channel. Figure \ref{tauto3mu}
shows the experimentally allowed regions for this channel in the
parameter space $\chi^\ell_{2,\mu\mu}$ \emph{vs}
$\chi^\ell_{2,\tau\mu}$. Also, we have set the theoretical restriction
coming from the UC. The region of parameters are consistent with
similar analysis in the context of the 2HDM-III\cite{Atwood:1996vj}.

The interval for $|\chi^\ell_{2,\tau\mu}|$ is determined by the upper
bound for the branching ratio. We have obtained a similar behavior to
the case $\tau\to\gamma\mu$, where we notice that for small values of
$|\chi^\ell_{2,\mu\mu}|$ the UC is more important than the limit
obtained from the experimental measurement. Thus, the LFV contribution
is suppressed by the texture given by the expressions
(\ref{eq:VSpinChi0}-\ref{eq:VSpinChi}). Comparing the interval of
values for $\chi^\ell_{2,\tau\mu}$ in Figure \ref{tautogmu} and
\ref{tauto3mu}, we can see that the UC is more restrictive for the
channel $\tau\to\gamma\mu$ than for $\tau\to 3 \mu$. Thus, in order to
test the UC, the process $\tau\to\gamma\mu$ is more important.

\subsubsection{The $B^0_s\to\mu^+\mu^-$ channel}
This channel is a rare decay because in the SM it is helicity
suppressed and it contains FCNC\cite{Buras:2012ru,Bobeth2014d}. Since
the observation of this decay\cite{Aaij:2012nna,Aaltonen:2013as} and
its compatibility with the SM
prediction\cite{CMSandLHCbCollaborations:2013pla}, it has been
used to test several
models\cite{Fleischer:2014jaa,Guadagnoli:2013mru}, from the 2HDM
\cite{Crivellin:2013wna} to Supersymmetry (See [\citen{Arbey:2012ax}]
and references therein) and also models with Spin 2 mediators
\cite{Fajfer:2018lix} and leptoquarks\cite{Smirnov:2018ske} to mention
some examples. A recent experimental study for this decay can be seen
in reference [\citen{Aaboud:2018mst}].

In general, the Wilson operators that contribute to leptonic decays of
pseudoscalar mesons at low energies consider whether axial, vector,
scalar and pseudoscalar operators \cite{Buchalla:1995vs}. In a general
2HDM, the NP contributions to these decays are only the axial and the
pseudoscalar when we restrict ourselves to tree
level\cite{Crivellin:2013wna}. The Wilson operators that parameterize
such effects are
\begin{equation}
  \label{eqn:operadores}
  \begin{split}
    \mathcal{O}_A^{q_fq_i} &=\left(\bar{q}_f\gamma_{\mu}P_Lq_i\right)\left(\bar{l}_B\gamma^{\mu}\gamma_5l_A\right),\\
    \mathcal{O^{'}}_A^{q_fq_i}&= \left(\bar{q}_f\gamma_{\mu}P_Rq_i\right)\left(\bar{l}_B\gamma^{\mu}\gamma_5l_A\right),\\
    \mathcal{O}_P^{q_fq_i}&= \left(\bar{q}_fP_Lq_i\right)\left(\bar{l}_B\gamma_5l_A\right),\\
    \mathcal{O^{'}}_P^{q_fq_i}&=
    \left(\bar{q}_fP_Rq_i\right)\left(\bar{l}_B\gamma_5l_A\right).
  \end{split}
\end{equation}
Thus, the effective Hamiltonian is given by
\begin{equation}
  \label{eqn:Heffc}
  \begin{split}
    \mathcal{H}_{eff}=-\frac{G_F^2M_W^2}{\pi^2}[C_V^{q_fq_i}O_V^{q_fq_i}+C_A^{q_fq_i}O_A^{q_fq_i}+C_S^{q_fq_i}O_S^{q_fq_i}+C_P^{q_fq_i}O_P^{q_fq_i}\\+
    \text{primed}] + H.c..
  \end{split}
\end{equation}
The branching ratio for this decay is reduced to
\begin{equation}
  \label{eqn:BR^{Exp}}
  \begin{split}
    \text{Br}[P^0_{\bar{q}_f,q_i}\to
    l_A^+l_B^-]&=\frac{G_F^4M_W^4}{32\pi^5}f(x_A^2,x_B^2)M_{P^0}f_{P^0}^2(m_{l_A}+m_{l_B})^2\tau_{P^0}
    [1-(x_A-x_B)^2] \\ & \left| \frac{M_{P^0}^2\left(
          C_P^{q_fq_i}-C_p^{'q_fq_i}\right)}{\left(m_{q_f}+m_{q_i}\right)\left(m_{l_A}
          + m_{l_B}\right)}-\left(C_A^{q_fq_i}-C_A^{'q_fq_i}
      \right)\right|^2,
  \end{split}
\end{equation}  
where $f(x_i,x_j)$ is the kinematical function defined by
\begin{eqnarray}
  f(x_i,x_j)=\sqrt{1-2\left(x_i+x_j\right)+\left(x_i-x_j\right)^2},
\end{eqnarray}
where $x_i=\frac{m_{l_i}}{M_{P^0}}$.  The SM contributions to the
branching ratio are contained in the Wilson coefficient $C_A$. The SM
contribution for this channel has been calculated for the pseudoscalar
meson $B$ in [\citen{Bobeth:2013tba,Bobeth:2013uxa,Bobeth:2001sq}]. In
general, the axial contributions for a pseudoscalar meson decay can be
written as \cite{Crivellin:2013wna}
\begin{equation}
  C_A^{q_fq_i}-C_A^{'q_fq_i}=-V_{tq_i}^*V_{tq_f}Y(\frac{m_t^2}{M_W^2})-V_{cq_i}^*V_{cq_f}Y(\frac{m_c^2}{M_W^2}),
\end{equation}
where $V_{q_kq_l}$ are the CKM matrix elements and
$Y(x)=\eta_Y Y_0(x)$ is the widely known Inami-Lim
function\cite{Inami:1980fz}, that includes the NLO QCD
effects\cite{Buras:2012ru} with $\eta_Y=1.0113$.  The NP tree level
contribution is parameterized by the Wilson coefficients
\begin{equation}\label{Wilson_PCoeffcient}
  C_P-C'_P=\frac{-i}{m_A^2}\Gamma_{A^0q_iq_f}\Gamma_{A^0\ell_A \ell_B},
\end{equation}
where the couplings are given by (\ref{couplingsGammaA}) and

\begin{equation}
  \label{couplingsAqiqf}
  \Gamma_{A^0q_fq_i}=\frac{i }{2}\left[-m_i \tan\beta \delta_{fi}+\frac{\sqrt{m_{q_i}m_{q_f}}}{\sqrt{2}\cos\beta} \tilde \chi_{2,q_fq_i}^d\right].
\end{equation}

As a result, we obtain the branching ratio as a function of the
pseudoscalar mass $M_{A^0}$, $\tan\beta$, the mixing angle $\alpha$
and the parameters $c^V,c^0,c_3^V$.

\subsubsection{$B_s^0-B_s^0$ mixing}
In the quark sector, the most important restriction to NP that
generates FCNCs comes from the mixing of pseudoscalar neutral
mesons. Actually, the $K^0-\bar K^0$ highly restricts the contribution
that might come from the interaction with the first generation of
quarks. In the case when there exist mixing only between second and
third generation, the only observable that might be deviated from the
SM prediction is $\Delta M_{B^0_s}$. In our previous
work\cite{Carrillo-Monteverde2019}, we have shown that it is possible to
find a bound on the non standard scalar contribution to this
observable assuming that the experimental measurement contains two
terms of the form
\begin{equation}
  \label{eq:mass_spliting}
  \Delta M_{q}=\Delta M^\text{SM}_{q}+\Delta M^{NP}_{q}.
\end{equation}
where $\Delta M^\text{SM}_{q}$ is the SM prediction and
$\Delta M_q^\text{NP}$ can be parameterized by
\begin{equation}
  \label{eq:deltamass_NP}
  \Delta M_q^\text{NP}(\mu)=|C_\text{NP}|\frac{M_{B_q}^2f_{B_q}^2}{\left[m_b(\mu)+m_q(\mu)\right]^2}\sigma(\mu),
\end{equation}
where we have define
$\mathcal
\sigma(\mu)=|-\frac{5}{24}B_2(\mu)+\frac{1}{24}B_3(\mu)+\frac{1}{24}B_4(\mu)+\frac{1}{12}B_5(\mu)|$
and $B_i(\mu)$ (for $i=2,3,4,5$) are the renormalized bag parameters
at the scale $\mu$
($\mu=m_b$)\cite{Becirevic:2001jj,Becirevic:2001xt,Huang:2004pk}. The
coefficient $|C_\text{NP}|$ is calculated at the low energy limit of
the generic 2HDM and it has the form
\begin{equation}
  \label{eq:coefficient_NP}
  |C_\text{NP}|=\frac{1+\tan^2\beta}{2M_A^2}|\widetilde Y_{2,ij}^q|,
\end{equation}
where $\widetilde Y_{2,ij}^q$ is the corresponding Yukawa matrix
element in the mass basis.  Thus, the upper bound for
$|\tilde\chi^d_{2,bq}|$ is given by
\begin{equation}
  \label{eq:up_bound}
  |\tilde \chi^d_{2,bq}|\leq\left(\frac{2M_A^2}{1+\tan^2\beta}\right)\frac{v(m_b+m_q)^2}{\sqrt{m_b m_q}M_{B_q}^2f_{B_q}^2}\frac{E_{B_q}}{\mathcal \sigma(m_b)},
\end{equation}
where we have defined the parameter
\begin{equation}
  E_{B_q}=|\Delta M^\text{SM}-\Delta M^\text{Exp}|-\sqrt{\delta^2(\Delta M^\text{SM})+\delta^2(\Delta M^\text{Exp})}.
\end{equation}

\subsection{Numerical analysis}\label{sec:numerical}

In order to find the combination of parameters that is compatible with
the UC, that is, the values of $c^V$, $c_0^V$ and $c_3^V$ that render
values for the branching ratios of $\tau\to\gamma\mu$, $\tau\to 3\mu$
and $B^0_s\to\mu\bar\mu$ restricted by the narrow window for NP given
by the restriction of $\Delta M_{B^0_s}$, we built level curves in
the free parameters space with the following choice of parameters:
\begin{itemize}
\item The lightest scalar $h^0$ of the model was chosen as the Higgs
  boson of the SM, that is, $M_{h^0}=125.6$ GeV.
\item We are interested only in scenarios with relatively light non
  standard neutral scalars and pseudoscalars, thus we have set
  $M_{A^0}=300$ GeV and $M_{H^0}=300$ GeV.
\item The mixing angle was chosen as $\alpha=0$ without loss of
  generality.
\item For the free parameters in the texture, the phase has been set
  as $\theta_V=0$ and the intervals for $c^V$, $c_0^V$ and $c_3^V$ are
  determined by the phenomenology of the 4 selected observables.
\end{itemize}
\begin{figure}[ht]\label{UniversalityBound}
  \begin{center}
    \includegraphics[scale=0.45]{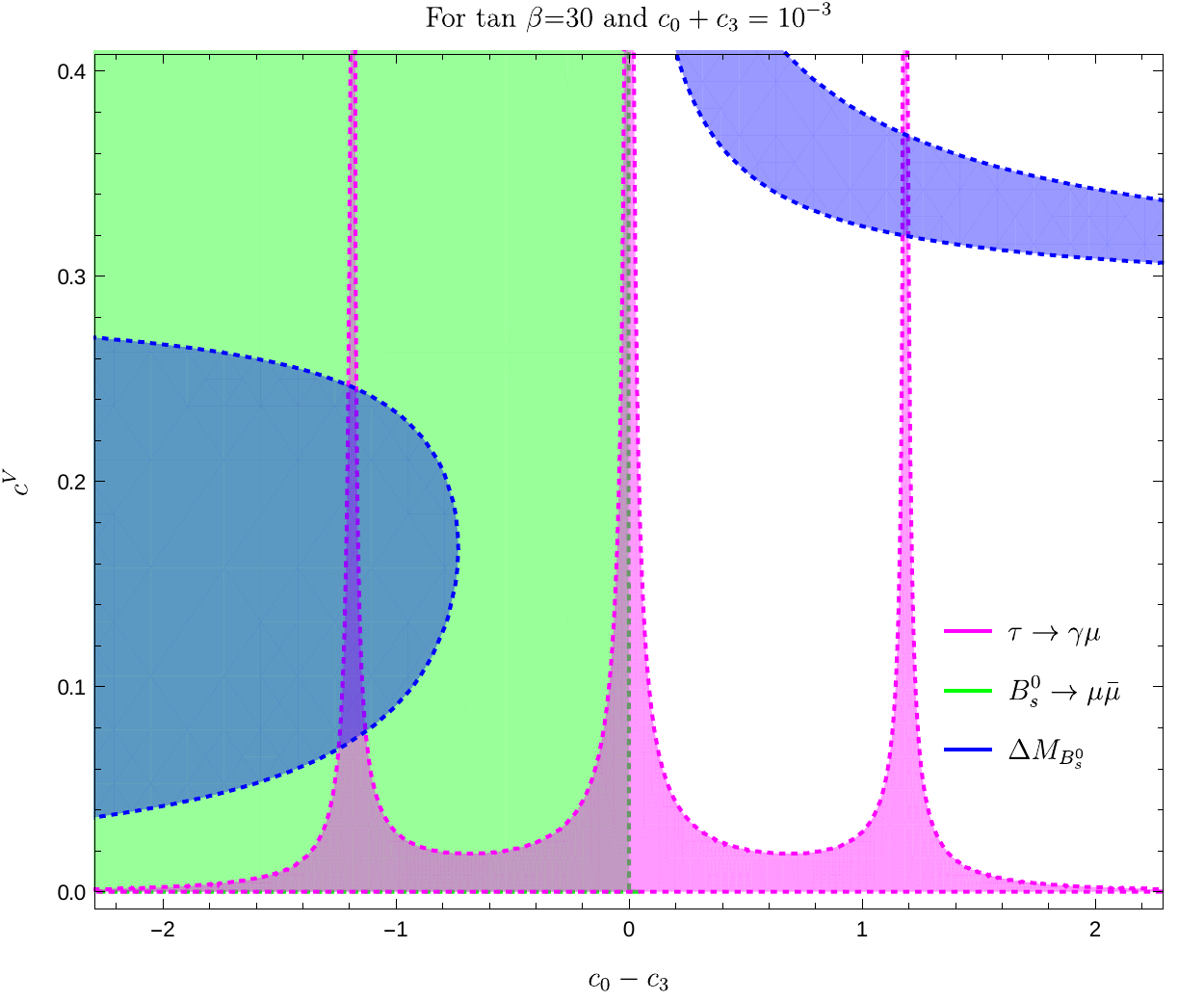}
    \includegraphics[scale=0.45]{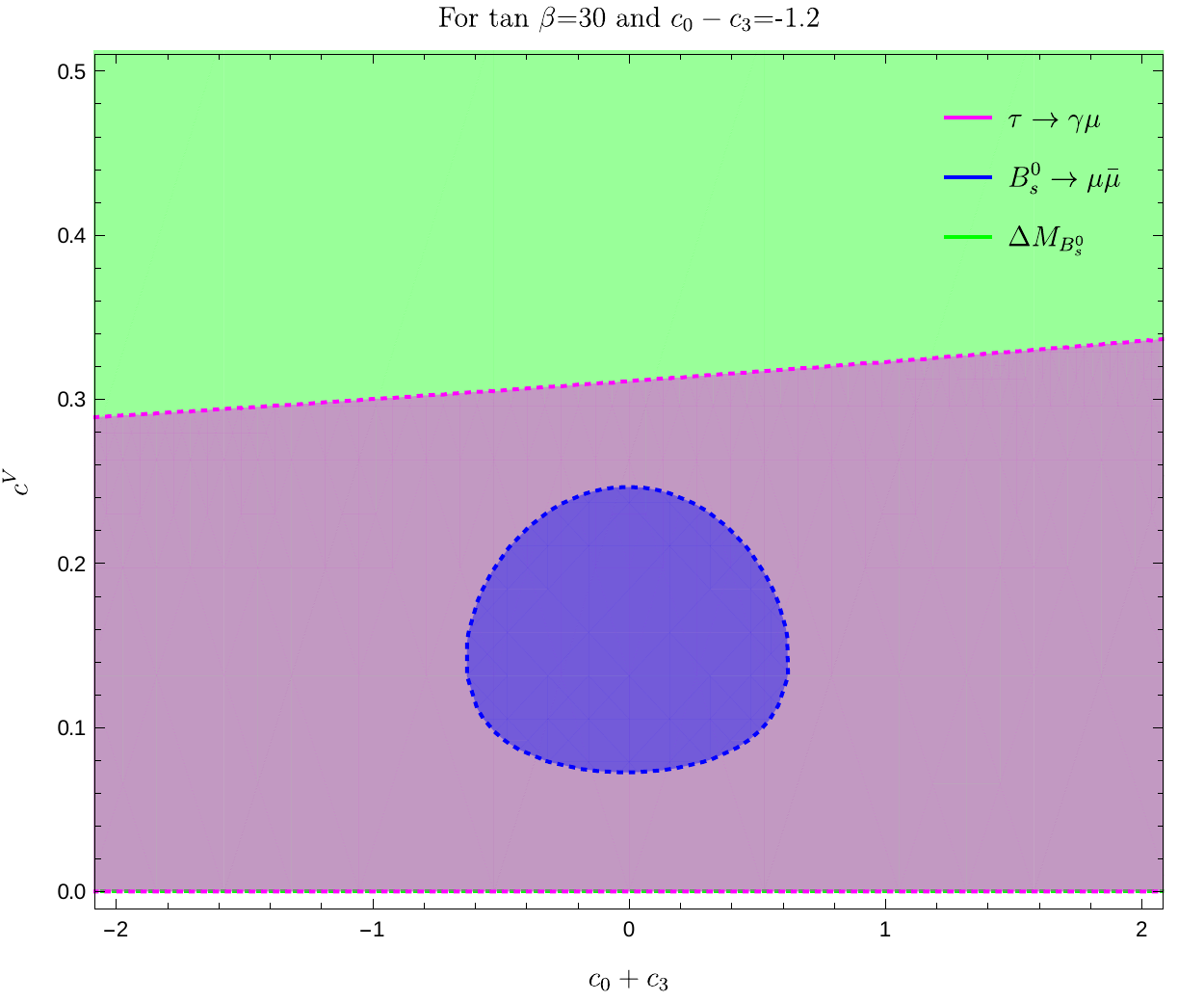}\\
    \includegraphics[scale=0.45]{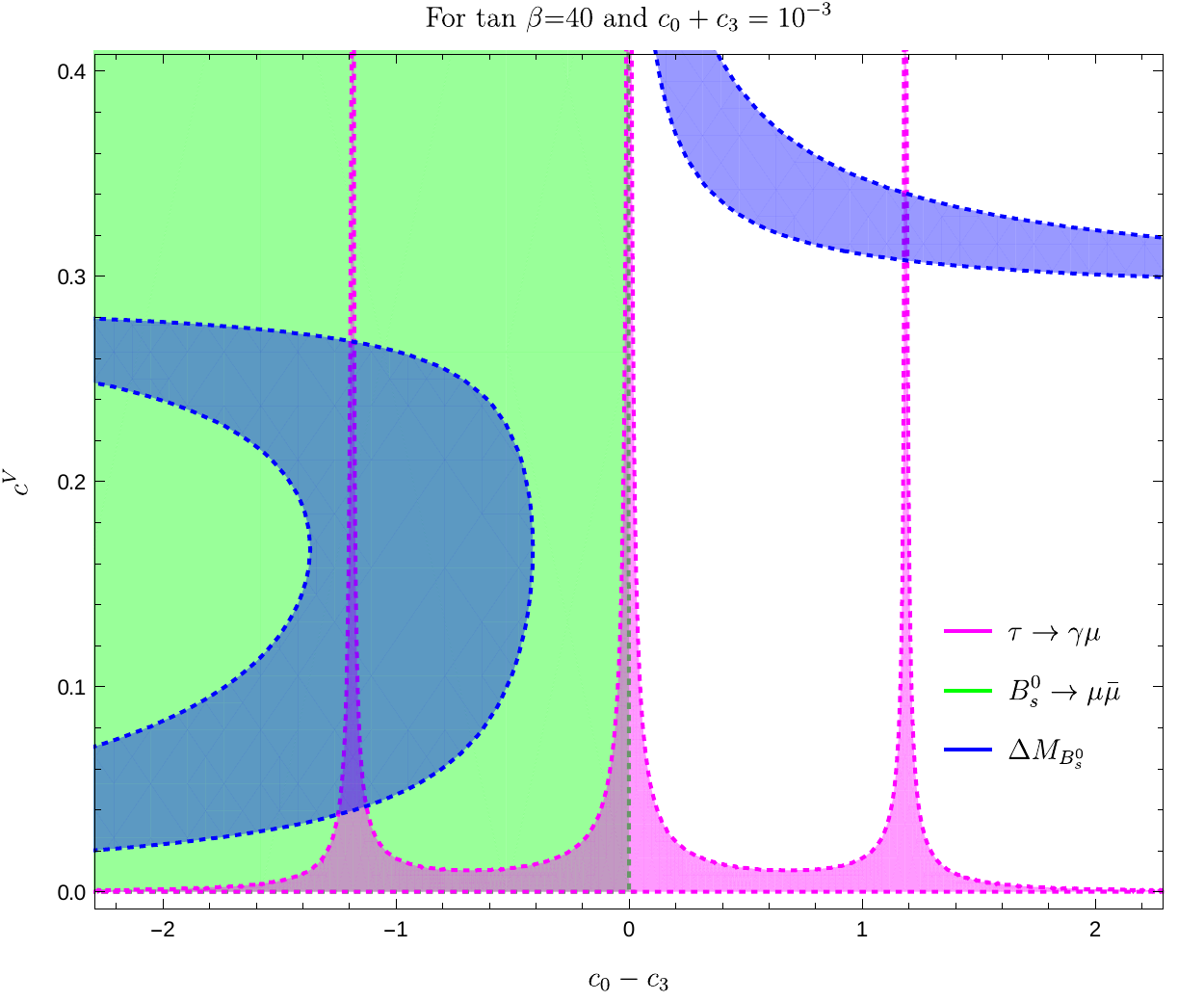}
    \includegraphics[scale=0.45]{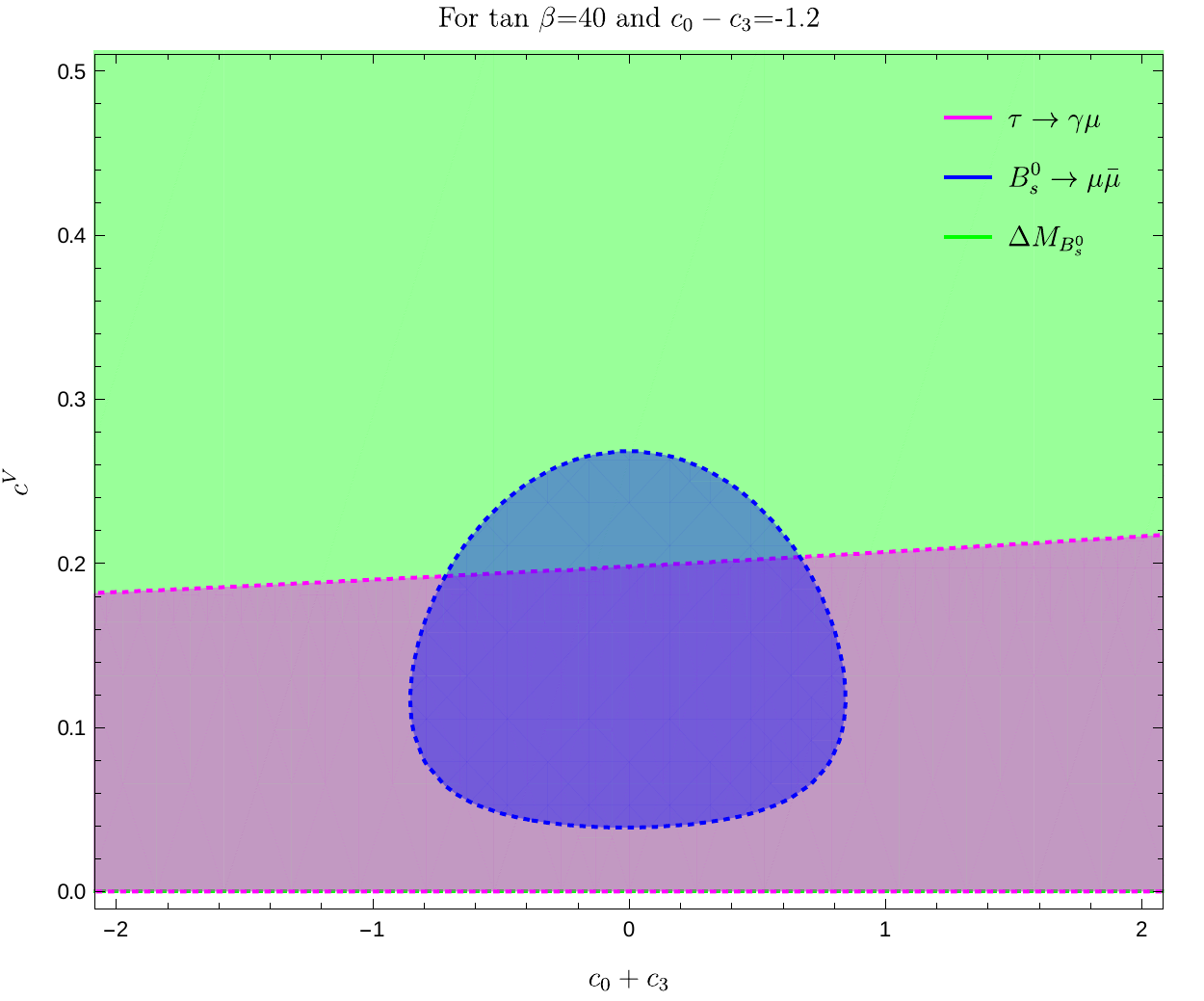}\\
    \includegraphics[scale=0.45]{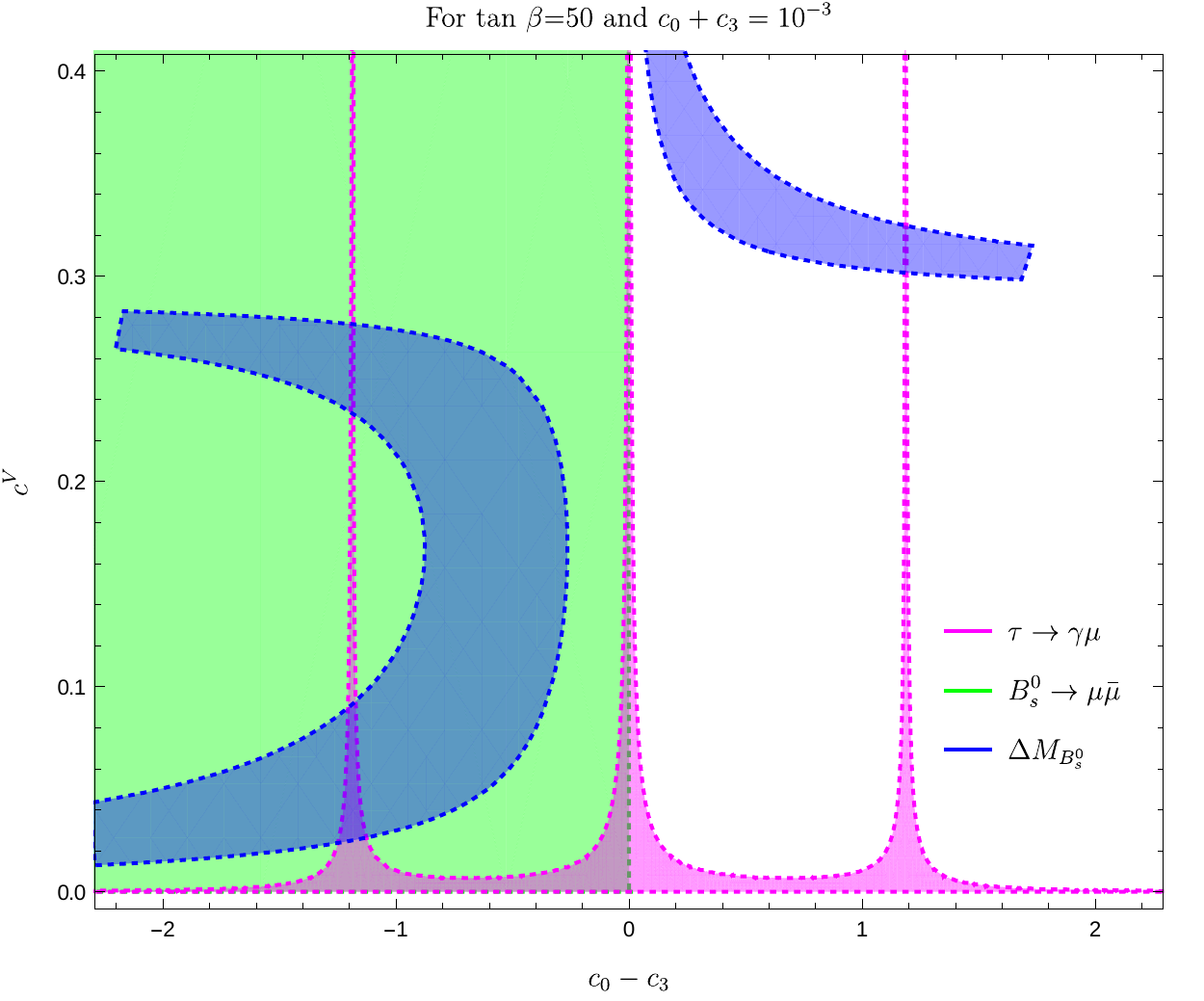}
    \includegraphics[scale=0.45]{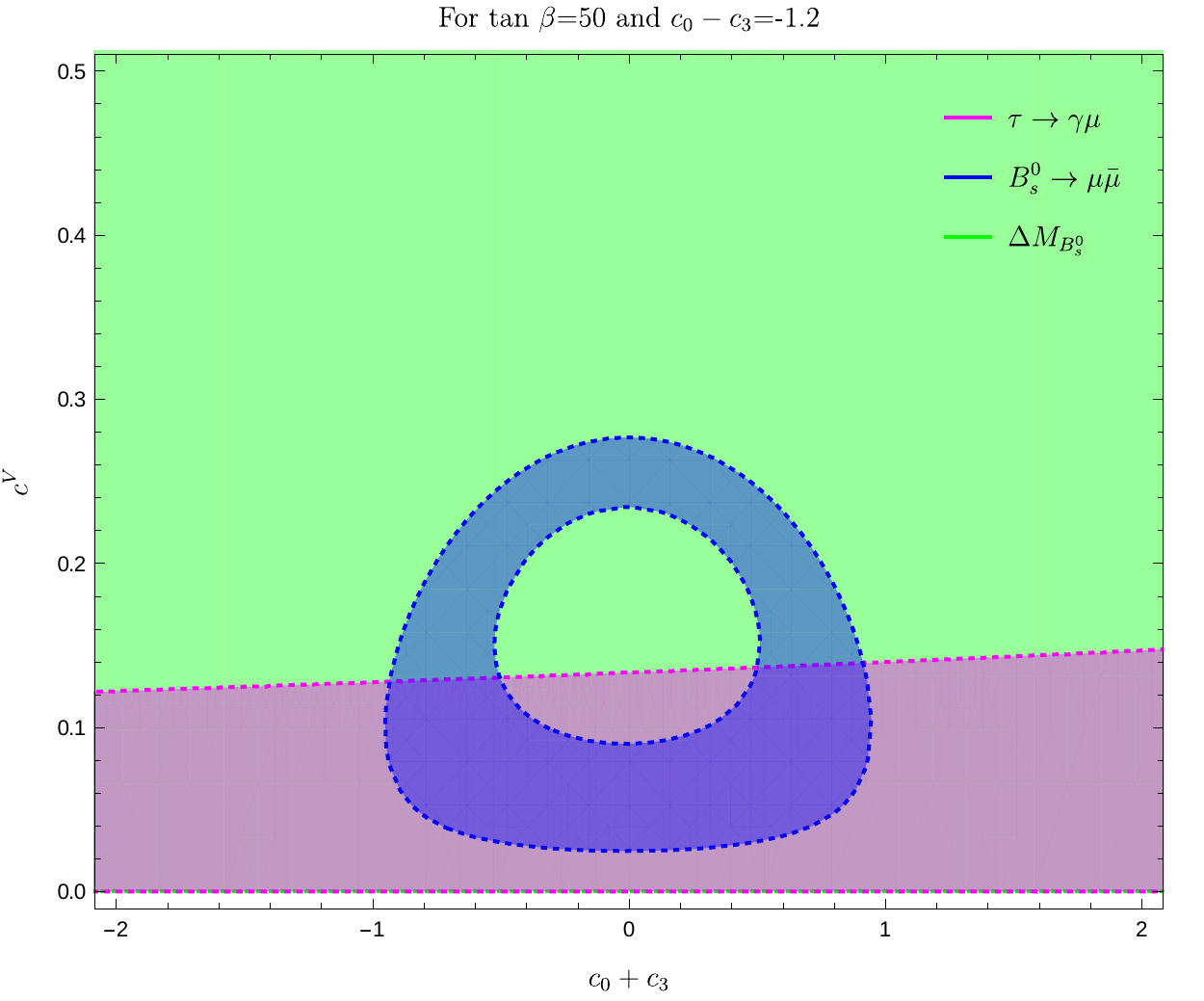}\\
  \end{center}
  \caption{Here, we show the allowed regions of parameters of the
    transformation (\ref{eq:flavor_transformation}) that fulfills the
    experimental bounds coming from the channels $\tau\to\gamma\mu$,
    $\tau\to 3\mu$ and $B^0_s\to\mu\bar\mu$. Also, it is included the
    restriction coming from the measurement of $\Delta M_{B^0_s}$. The
    region for $\tau\to 3\mu$ is wider and includes the other
    processes, thus we focus on the regions shown here. On the left, it
    is shown the narrow region of values for $c_0^V-c_3^V$ and $c^V$
    for different values of $\tan\beta$ with
    $M_{h^0}=125.6$ GeV, $M_{H^0}=300$ GeV and $M_{A^0}=300$ GeV. The
    sum of parameters was fixed as $c_0^V+c_3^V=10^{-3}$. The right-hand
    side plots were generated fixing $(c_0^V-c_3^V)\simeq -1.2$. The
    area where the allowed region are superposed gives the interval of
    values for $c^V$, $c_0^V-c_3^V$ and $c_0^V+c_3^V$ where the
    UC is valid. The region is sensitive to
    $\tan\beta$ and it disappears at $\tan\beta\gtrsim 80$.}
\end{figure}

In Figure \ref{UniversalityBound}, we look for the region of
parameters where UC bound is valid using the 4 channels that contain
NP contribution from the V-Spin texture. The region given by $\tau\to3\mu$ is wider than those from the rest of the processes, thus in the following plot we concentrate our the attention
to the region that come from the bounds on $\Delta M_{B^0_s}$,
$\text{BR}(\tau\to\gamma\mu)$ and $\text{BR}(B_s^0\to\mu\bar\mu)$. The
main result of this work is reviewed in Figure
\ref{UniversalityBound}, where we have plotted the experimental
allowed region described by the parameters $c^V$, $c^V_0-c_3^V$ and
$c^V_0+c_3^V$. The left-hand side plots were generated fixing
$c_0^V+c_3^V\simeq\mathcal{O}(10^{-3})$ . As we can see from the plots
on the right side, the only place where there is a superposition of
parameters is when $c_0^V-c_3^V\simeq -1.2$. This region is very
sensitive to the values of $\tan\beta$ and in our analysis we found that
for $\tan\beta\gtrsim 80$ the UC coming from the V-Spin cannot
describe simultaneously the updated experimental bounds for the three
channels with the chosen scalar masses in this analysis.

\section{Conclusions}\label{sec:conc}
At this point, we want to review this work. We have used the V-Spin
case of a PA-2HDM model in order to describe possible FCNC and LFV at
tree level as the result of the interchange of non-standard scalars.
The V-Spin case gives a texture for the second Yukawa matrix where
only the second and the third generation of fermions are mixed. In
consequence, only the channels $\tau\to\gamma\mu$, $\tau\to3\mu$ in
the leptonic sector and $B_s^0\to\mu\bar\mu$ with $\Delta M_{B^0_s}$
for quarks might have effects of NP that could be measured in current
experiments.

Also, we have assumed that the texture for the second Yukawa matrix is
universal, that is, that the same dimensionless parameters enter the
quark and lepton couplings, and the difference in mixings comes from
the fermion mass hierarchy. This assumption introduces a relation
between the Yukawa matrix elements for different sectors; we have
called this relation the Universality Constraint. For this reason, the
phenomenological bounds in one sector impacts the predictions for
other sectors. In this work, we have used the Cheng and Sher
parametrization of the Yukawa matrix elements in order to facilitate a
comparison with past analysis on the generic 2HDM and to express the
UC in the parameter spaces $\chi^\ell_{2,\mu\mu}$ vs
$\chi^\ell_{2,\tau\mu}$ and $\chi^\ell_{2,\mu\mu}$ vs
$\chi^\ell_{2,\tau\tau}$. This was shown in the figures \ref{tauto3mu} and \ref{tautogmu}. These regions were built with the LFV
processes $\tau\to\gamma\mu$ and $\tau\to 3\mu$ respectively.

Under the universal texture scheme, the restrictive measurements of
B-meson decays require the suppression of LFV effects in
$\tau\to\gamma\mu$ and $\tau\to3\mu$. The allowed regions were shown
in Figures \ref{tautogmu} and \ref{tauto3mu} where the superposition
region is such that for small values of $\chi^\ell_{2,\tau\tau}$ and
$\chi^\ell_{2,\mu\mu}$, the main restriction on $\chi^\ell_{2,\tau\mu}$
comes from the UC. For large values, the experimental bound is
stronger for this parameter.

We have found a region in the parameter space where the UC is
phenomenologically viable, testing the universality hypothesis using
the channels of Table \ref{channelsVSpin}. This requires that the
dimensionless parameters satisfy $(c_0^V-c_3^V)\sim -1.2$ and
$(c_0^V+c_3^V)\sim\mathcal{O}(10^{-3})$.  The channel
$B_s^0\to\mu\bar\mu$ gives the stronger constraint, and updates to
experimental precision of this channel could challenge the UC. For
relatively light masses $M_{h^0}=125.6$ GeV, $M_{H^0}=300$ GeV and
$M_{A^0}=300$ GeV, the allowed region disappear at values of
$\tan\beta\gtrsim 80$.
\bibliography{bibliography}
\end{document}